\documentclass[aps,prb,reprint,superscriptaddress,showpacs]{revtex4-1}
\usepackage{graphicx}
\usepackage{dcolumn}
\usepackage{bm}
\usepackage{epstopdf}
\usepackage{hyperref}
\usepackage[utf8]{inputenc}
\usepackage[T1]{fontenc}
\usepackage{lmodern}
\usepackage{amsmath}
\usepackage{amssymb}
\usepackage{booktabs}
\usepackage{float}

\begin{document}


\title{Absence of magnetic long range order in Y$_{2}$CrSbO$_{7}$: bond-disorder induced magnetic frustration in a ferromagnetic pyrochlore}


\author{L. Shen}
\affiliation{School of Physics and Astronomy, University of Birmingham, Birmingham B15 2TT, United Kingdom}

\author{C. Greaves}
\affiliation{School of Chemistry, University of Birmingham, Birmingham B15 2TT, United Kingdom}

\author{R. Riyat}
\affiliation{School of Physics and Astronomy, University of Birmingham, Birmingham B15 2TT, United Kingdom}

\author{T. C. Hansen}
\affiliation{Institut Laue-Langevin, BP 156, 38042 Grenoble Cedex 9, France}

\author{E. Blackburn}
\affiliation{School of Physics and Astronomy, University of Birmingham, Birmingham B15 2TT, United Kingdom}



\begin{abstract}
The consequences of nonmagnetic-ion dilution for the pyrochlore family Y$_{2}$(\textit{M}$_{1-x}$\textit{N}$_{x}$)$_{2}$O$_{7}$ (\textit{M}\ =\ magnetic ion, \textit{N}\ =\ nonmagnetic ion) have been investigated.\ As a first step, we experimentally examine the magnetic properties of Y$_{2}$CrSbO$_{7}$ (\textit{x}\ =\ 0.5), in which the magnetic sites (Cr$^{3+}$) are percolative.\ Although the effective Cr-Cr spin exchange is ferromagnetic, as evidenced by a positive Curie-Weiss temperature, $\Theta_\mathrm{{CW}}$\ =\ 20.1(6)\,K, our high-resolution neutron powder diffraction measurements detect no sign of magnetic long range order down to 2\,K. In order to understand our observations, we performed numerical simulations to study the bond-disorder introduced by the ionic size mismatch between \textit{M} and \textit{N}. Based on these simulations, bond-disorder ($x_{b}$\ $\simeq$\ 0.23) percolates well ahead of site-disorder ($x_{s}$\ $\simeq$\ 0.61). This model successfully reproduces the critical region (0.2\ <\ \textit{x}\ <\ 0.25) for the N\'eel to spin glass phase transition in Zn(Cr$_{1-x}$Ga$_{x}$)$_{2}$O$_{4}$,  where the Cr/Ga-sublattice forms the same corner-sharing tetrahedral network as the \textit{M}/\textit{N}-sublattice in Y$_{2}$(\textit{M}$_{1-x}$\textit{N}$_{x}$)$_{2}$O$_{7}$, and the rapid drop in magnetically ordered moment in the N\'eel phase [Lee \textit{et al}, Phys. Rev. B 77, 014405 (2008)].\ Our study stresses the nonnegligible role of bond-disorder on magnetic frustration, even in ferromagnets.
\end{abstract}

\maketitle
\section{Introduction}
Magnetic frustration, which often leads to interesting spin structures, refers to systems where the total free energy cannot be minimized by optimizing the interaction energy between each pair of spins\,\cite{Diep}.\ Magnetic interactions can be frustrated by geometry. For example, magnetic long range order is prohibited for the Heisenberg antiferromagnet on a triangular (or tetrahedral) lattice\,\cite{Gardner}. The corresponding magnetic ground state, named spin liquid, is highly degenerate\,\cite{Moessner,Moessner2,Canals}. In addition to geometry, the competition between different types of magnetic interactions can also lead to magnetic frustration. The Ising rare-earth pyrochlores \textit{R}$_{2}$Ti$_{2}$O$_{7}$ (\textit{R}\ =\ Ho,\,Dy), in which the \textit{R}-sublattice forms a corner-sharing tetrahedral network, develop a novel two-in/two-out spin ice structure due to the competing exchange and dipole-dipole interactions\,\cite{Gardner, Harris, Bramwell}. Strikingly, the excited quasiparticles of a spin ice are found to resemble the behaviour of magnetic monopoles\,\cite{Morris, Castelnovo}.

The effect of disorder has been widely investigated in magnetic materials and disorder is commonly used to generate spin glasses\,\cite{Binder}. In general, a spin glass (SG) state prevails in systems dominated by randomness and frustration, which can be realized by either site- or bond-disorder.

Site-disorder arises when ions with different magnetic properties may be found randomly distributed on the same crystallographic sites, and is a very effective way to frustrate the Ruderman-Kittel-Kasuya-Yosida (RKKY) interaction.\ SG alloys such as Cu-Mn, Au-Mn and Au-Fe belong to this category\,\cite{Nagata,Morgownik}.\ Moreover, SG can also be induced by diluting magnetic sites using nonmagnetic ions to pass the site-disorder percolation threshold ($x_{s}$), as in\ Eu$_{1-x}$Sr$_{x}$S ($x_{s}$\ $\approx$\ 0.136)\,\cite{Binder,Maletta}.

Bond-disorder arises due to randomization of bond-length. Recent theoretical advances\,\cite{Saunders,Shinaoka} strongly suggest that bond-disorder is essential to generate a SG state in magnetic pyrochlores and spinels such as Y$_{2}$Mo$_{2}$O$_{7}$\,\cite{Gardner,Gingras, Miyoshi, Greedan} and Zn(Cr$_{1-x}$Ga$_{x}$)$_{2}$O$_{4}$\,\cite{Ratcliff,Fiorani,Lee}.\ In the latter compound, the SG is not related to the site-disorder since the onset composition of SG, 0.2\ <\ \textit{x}\ <\ 0.25\,\cite{Lee}, is well below the percolation threshold of the nonmagnetic Ga$^{3+}$-sites, $x_{s}$\ $\approx$\ 0.61\,\cite{Henley}.

To the best of our knowledge, whether bond-disorder alone can lead to magnetic frustration remains to be unveiled. Although bond-disorder is decisive in Y$_{2}$Mo$_{2}$O$_{7}$ and Zn(Cr$_{1-x}$Ga$_{x}$)$_{2}$O$_{4}$\,\cite{Saunders,Shinaoka}, its influence on magnetic frustration is not clear due to the coexisting geometric frustration in these materials.\ Theoretically, it is argued that the weak bond-disorder acts as a perturbation to partially lift the degeneracy of a spin liquid\,\cite{Saunders}. From this point of view, bond-disorder does not facilitate magnetic frustration in systems composed of antiferromagnetically coupled spins. In addition, neither of the theories mentioned above could reproduce the critical region (0.2\ <\ \textit{x}\ <\ 0.25) for the N\'eel to SG phase transition in Zn(Cr$_{1-x}$Ga$_{x}$)$_{2}$O$_{4}$\,\cite{Lee}.  
\begin{figure*}
	\centering
	\includegraphics[width=0.85\textwidth]{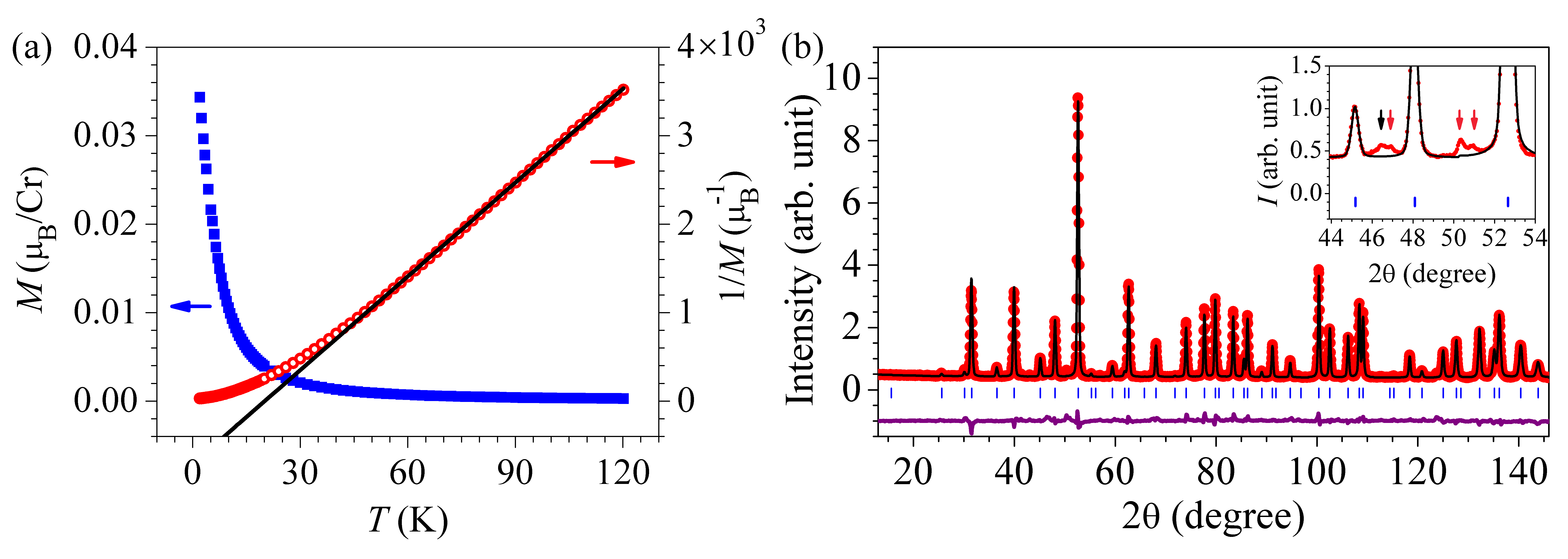}
	\caption{(a) Temperature dependences of magnetization (left axis, blue solid squares) and inverse magnetization (right axis, red open circles) of Y$_{2}$CrSbO$_{7}$ measured at $\mu_\mathrm{{0}}H$\,=\,0.01\,T. The black solid line is a Curie-Weiss fit to the linear part of the 1/\textit{M}-\textit{T} curve in the paramagnetic region. (b) HRNPD pattern (red solids) of Y$_{2}$CrSbO$_{7}$ at \textit{T}\,=\,2.0\,K, \textit{B}\,=\,0\,T. Calculated pattern (black line), nuclear Bragg positions (blue vertical line) and difference (purple line) are also displayed. (inset) Enlarged version of a selected angle region. Additional peaks from YCrO$_{3}$ (red arrows) and the vanadium sample can (black arrow) can be seen.}	
	\label{fig:1}
\end{figure*}

To demonstrate the exclusive influence of bond-disorder on magnetic frustration, or in other words, to avoid geometric frustration, we have studied a series of diluted ferromagnetic transition-metal (TM) pyrochlores,\ Y$_{2}$(\textit{M}$_{1-x}$\textit{N}$_{x}$)$_{2}$O$_{7}$ (\textit{M}\ =\ magnetic TM ion, \textit{N}\ =\ nonmagnetic ion), where the yttrium-sites are nonmagnetic and bond-disorder is introduced by the ionic size mismatch between \textit{M} and \textit{N}. According to Ref.~\onlinecite{Henley}, we expect \textit{N}-sites to percolate at $x_{s}$\ $\approx$\ 0.61. Specifically, we have employed Y$_{2}$Mn$^{4+}_{2}$O$_{7}$\,\cite{Shimakawa} (\textit{x}\ =\ 0) as the bond ordered start compound and Y$_{2}$(Cr$^{3+}_{1-x}$Ga$^{3+}_{x-0.5}$Sb$^{5+}_{0.5}$)$_{2}$O$_{7}$ (0.5 $\leqslant$\,\textit{x}\,$\leqslant$\,0.9) as the bond disordered compounds. The magnetic TM ions in these systems (Mn$^{4+}$ and Cr$^{3+}$) share the same electronic configuration (3$d^{3}$). Experimentally, we have performed magnetization and high-resolution neutron powder diffraction (HRNPD) measurements on Y$_{2}$(Cr$_{1-x}$Ga$_{x-0.5}$Sb$_{0.5}$)$_{2}$O$_{7}$ (0.5 $\leqslant$\,\textit{x}\,$\leqslant$\,0.9). In Y$_{2}$CrSbO$_{7}$ (\textit{x}\ =\ 0.5\ <\ $x_{s}$), we find no evidence of zero-field magnetic long range order down to 1.8\,K despite having a positive Curie-Weiss temperature $\Theta_\mathrm{{CW}}$\ =\ 20.1(6)\,K. Since Cr-sites are percolative in Y$_{2}$CrSbO$_{7}$\,\cite{Henley}, site-disorder cannot be the driving mechanism of the observed high magnetic frustration.\ We have also carried out comprehensive numerical simulations to study the percolation processes of various nonmagnetic clusters, including bond-disorder, site-disorder, and the intermediate types in between (see Section\ \ref{III}).\ Based on these simulations, bond-disorder percolates at $x_{b}$\ =\ 0.23(1) on a pyrochlore lattice, pointing to the percolative bond-disorder in Y$_{2}$CrSbO$_{7}$ (\textit{x}\ =\ 0.5). Our model also explains why the N\'eel to SG phase transition in Zn(Cr$_{1-x}$Ga$_{x}$)$_{2}$O$_{4}$ happens between \textit{x}\ =\ 0.2 and 0.25, as well as its rapid drop in magnetically ordered moment in the N\'eel phase upon Ga-substitution\,\cite{Lee}.\ The nonnegligible role of bond-disorder in the zero-field magnetic frustration in Y$_{2}$CrSbO$_{7}$ is further supported by recovering the magnetic long range order at high magnetic fields. 
\section{Experiments}
Polycrystalline samples of Y$_{2}$(Cr$_{1-x}$Ga$_{x-0.5}$Sb$_{0.5}$)$_{2}$O$_{7}$ (0.5 $\leqslant$\,\textit{x}\,$\leqslant$\,0.9) were synthesized by the solid-state reaction method in three steps\,\cite{Whitaker}. First of all, GaSbO$_{4}$ (CrSbO$_{4}$) powders were prepared by heating Ga$_{2}$O$_{3}$ (Cr$_{2}$O$_{3}$) (3N) and Sb$_{2}$O$_{3}$ (3N, 5\,\% excess to compensate the volatilization) for 3 days at 640\,$^{\circ}$C, and then 5 days at 1200\,$^{\circ}$C with several intermediate regrindings. The intermediate temperature (640\,$^{\circ}$C) is to transform Sb$_{2}$O$_{3}$ into Sb$_{2}$O$_{4}$. To prepare Y$_{2}$GaSbO$_{7}$ (Y$_{2}$CrSbO$_{7}$), a stoichiometric mixture (1\,:\,1) of GaSbO$_{4}$ (CrSbO$_{4}$) and Y$_{2}$O$_{3}$\,(4N) were heated in air for 6 days at 1200\,$^{\circ}$C with several intermediate regrindings as well.\ Finally, Y$_{2}$(Cr$_{1-x}$Ga$_{x-0.5}$Sb$_{0.5}$)$_{2}$O$_{7}$ was obtained by heating the stoichiometrically mixed Y$_{2}$GaSbO$_{7}$ and Y$_{2}$CrSbO$_{7}$ powders for 5 days at 1200\,$^{\circ}$C.

Magnetization data were recorded using a Magnetic Property Measurement System (MPMS, Quantum Design). X-ray powder diffraction measurements were performed using a Bruker D8 diffractometer (Cu K$\alpha$1,\,$\lambda$\,=\,1.5406\,\AA{}) at room temperature. HRNPD patterns were collected at the D2B powder diffractometer ($\lambda$\,=\,1.594\,\AA{}), equipped with a 5 T vertical cryomagnet, at the Institute Laue-Langevin (ILL) in Grenoble, France\ \cite{data}.\ For these measurements, about 8\,g of each powder sample were hydraulically pressed into a cylinder (height\ =\ 11\,mm, diameter\ =\ 13\,mm) to avoid any field-induced texture and then loaded into a vanadium container. Rietveld refinements were carried out using the FullProf package\,\cite{FullProf}.

An impurity phase, identified as YCrO$_{3}$, with a volume fraction of 3.4(2)\,\%, is necessary to match some very weak peaks in our HRNPD patterns (inset of Fig.\,\ref{fig:1}b). The onset of antiferromagnetism in YCrO$_{3}$ is responsible for a kink around 142\,K in our magnetization curves (data not shown here)\,\cite{Whitaker}. As a result, we only show magnetization data measured below 120\,K in this paper.

\section{Results and discussion}\label{III}
\begin{table*}
  \centering
  \caption{Structural parameters of Y$_{2}$Mn$_{2}$O$_{7}$ (from Ref.~\onlinecite{Subramanian}), Y$_{2}$CrSbO$_{7}$ and Y$_{2}$Cr$_{0.4}$Ga$_{0.6}$SbO$_{7}$, in which \textit{B} represents the atomic positions of Mn/Cr/Sb/Ga. The corresponding diffraction patterns were refined under space group \textit{F d -3 m} (\textit{a\,=\,b\,=\,c, $\alpha$\,=\,$\beta$\,=\,$\gamma$\,=}\,90$^{\circ}$). The only atomic position that needs to be refined is O2 (\textit{x}, 0.125, 0.125).}
  \label{tab:I}
  \begin{ruledtabular}
  \begin{tabular}{cccccccccc}
  &&&\multicolumn{4}{c}{\textit{B}$\mathrm{_{iso}}$\,(\AA$^{2}$)}&&&\\
  \cmidrule{4-7} 
  
&\textit{a}\,(\AA)&\textit{x}\,(O2)&Y&\textit{B}&O1&O2& \textit{B}\,-\,O2\,(\AA)&\textit{B}\,-\,\textit{B}\,(\AA)&\textit{B}\,-\,O2\,-\,\textit{B}\,($^{\circ}$)\\  
  \hline
\multicolumn{1}{c|}{Y$_{2}$Mn$_{2}$O$_{7}$\ (RT)}&9.902(1)&0.3274(8)&0.3(1)&0.1(1)&0.1(3)&0.2(1)&1.911(3)&3.5009(3)&132.7(5)\\  
\multicolumn{1}{c|}{Y$_{2}$CrSbO$_{7}$\ (300\,K)}&10.1620(1)&0.4178(1)&0.72(1)&0.44(2)&0.15(3)&0.45(1)&1.9810(6)&3.59282(3)&130.14(2) \\
\multicolumn{1}{c|}{Y$_{2}$CrSbO$_{7}$\ (2.0\,K)}&10.1523(7)&0.41793(8)&0.69(1)&0.34(1)&0.17(2)&0.439(8)&1.9787(3)&3.5894(2)&130.19(1)\\
\multicolumn{1}{c|}{Y$_{2}$Cr$_{0.4}$Ga$_{0.6}$SbO$_{7}$\ (2.0\,K)}&10.1508(8)&0.4182(1)&0.58(1)&0.51(2)&0.17(3)&0.37(1)&1.9774(5)&3.58885(2)&130.31(2)\\
  \end{tabular}
  \end{ruledtabular}
\end{table*}
We first discuss the bond ordered compound Y$_{2}$Mn$_{2}$O$_{7}$\ (\textit{x}\ =\ 0).\ The predominant Mn-Mn exchange is ferromagnetic, as evidenced by $\Theta_\mathrm{{CW}}$\ =\ 41(2)\,K\,\cite{Reimers}. The effective magnetic moment ($M_\mathrm{{eff}}$) deduced from fitting the magnetization versus temperature (\textit{M-T}) curve in the paramagnetic region is 3.84(2)\,$\mu_{B}$/Mn, indicating \textit{J}\ =\ \textit{S}\ =\ 3/2 for Mn$^{4+}$ (\textit{J} and \textit{S} are the total and spin angular momenta, respectively)\,\cite{Reimers}. The orbital quenching in Y$_{2}$Mn$_{2}$O$_{7}$ is also confirmed by its saturation moment measured at 5\,K:\ $M_\mathrm{{s}}$\ $\approx$\ 3.0\,$\mu_{B}$/Mn \,\cite{Shimakawa}. The magnetic ground state of Y$_{2}$Mn$_{2}$O$_{7}$ is very sample dependent. Shimakawa \textit{et al.} claim ferromagnetism in their sample based on the $\lambda$ heat capacity anomaly around 15\,K, below which the \textit{M-T} curve plateaus\,\cite{Shimakawa}. However, the $\lambda$ anomaly in heat capacity is not observed in the samples prepared by Reimers \textit{et al.}\,\cite{Reimers} and Greedan \textit{et al.}\,\cite{Greedan2}. Instead, their results strongly support a SG-like state at low temperatures. The strong sample dependence of magnetic properties might be related to the valence disorder in Y$_{2}$Mn$_{2}$O$_{7}$, where high pressure synthesis is required to stabilize Mn$^{4+}$\,\cite{Shimakawa,Reimers,Fujinaka}.

While Cr$^{4+}$-based pyrochlores also require high pressure synthesis\,\cite{Fujinaka},\ Cr$^{3+}$-based pyrochlores can be prepared at ambient pressure\,\cite{Whitaker}. To avoid the potential complication to the magnetic structure caused by valence disorder, we have chosen to study the Cr$^{3+}$-based pyrochlore Y$_{2}$CrSbO$_{7}$, where Sb$^{5+}$ is used to compensate the valence loss. Cr$^{3+}$ shares the same $3d^{3}$ electronic configuration as Mn$^{4+}$.\ Moreover, the lattice parameters of Y$_{2}$CrSbO$_{7}$ are analogous to those of Y$_{2}$Mn$_{2}$O$_{7}$\ (Table\,\ref{tab:I})\,\cite{Subramanian}. As a result, the magnetic interactions in the two systems are expected to be similar. Deviation from paramagnetism can be observed in Y$_{2}$CrSbO$_{7}$ at low temperatures, as revealed by the inverse magnetization versus temperature (1/\textit{M-T}) curve, (Fig.\,\ref{fig:1}a). A linear fit to the 1/\textit{M-T} curve in the paramagnetic region gives $\Theta_\mathrm{{CW}}$\ =\ 20.1(6)\,K and $M_\mathrm{{eff}}$\ =\ 3.99(1)\,$\mu_{B}$/Cr. Since $\Theta_\mathrm{{CW}}$ is proportional to number of magnetic sites per unit cell\,$\times$\,exchange strength\,\cite{Blundell}, the Cr-Cr and Mn-Mn exchange strengths are very similar to each other in Y$_{2}$CrSbO$_{7}$ and Y$_{2}$Mn$_{2}$O$_{7}$.\ The \textit{M-T} curve is also displayed in Fig.\,\ref{fig:1}a.\ Surprisingly, no ferromagnetism can be observed down to 1.8\,K for Y$_{2}$CrSbO$_{7}$. The absence of magnetic long range order in Y$_{2}$CrSbO$_{7}$ is further confirmed by our HRNPD pattern at 2\,K, which can be refined by a single crystallographic phase (Fig.\,\ref{fig:1}b).

\begin{figure*}
	\centering
	\includegraphics[width=0.85\textwidth]{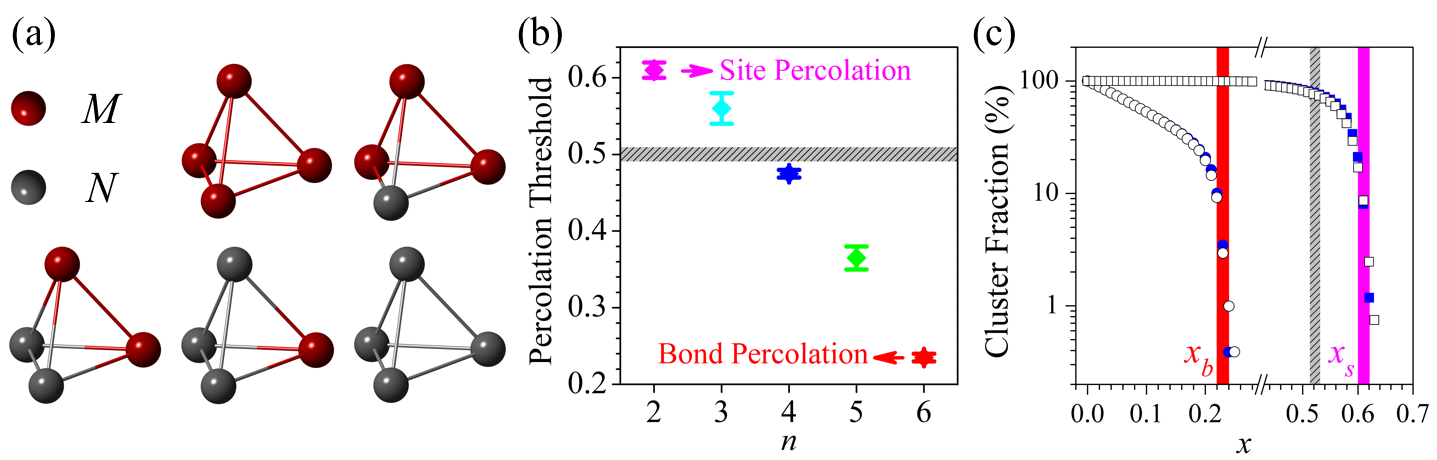}
	\caption{(a) Five possible configurations of a \textit{M}/\textit{N}-tetrahedron. (b) The percolation thresholds produced by our simulations on a \textit{D}\,$\times$\,\textit{D}\,$\times$\,\textit{D} pyrochlore lattice with \textit{s} sampling times of bond-disorder (red, \textit{D}\ =\ 64 and \textit{s}\ =\ 100), Cr-clusters with at least five (green, \textit{D}\ =\ 64 and \textit{s}\ =\ 16), four (blue, \textit{D}\ =\ 48 and \textit{s}\ =\ 16), three (cyan, \textit{D}\ =\ 48 and \textit{s}\ =\ 16) nearest neighbour Cr-sites, and site-disorder (magenta, \textit{D}\ =\ 64 and \textit{s}\ =\ 50). (c) The evolution of the fraction of percolative bond- (circles) and site- (squares) ordered clusters as a function of the nonmagnetic fraction \textit{x} (blue closed:\ \textit{D}\ =\ 48, black open:\ \textit{D}\ =\ 64). The red (magenta) area marks the percolation region of bond- (site-) disorder. The grey hatched areas in (b) and (c) mark the position of Y$_{2}$CrSbO$_{7}$ with $\Delta\,x$\ =\ 0.01.}	
	\label{fig:2}
\end{figure*}

\begin{figure}
	\centering
	\includegraphics[width=0.48\textwidth]{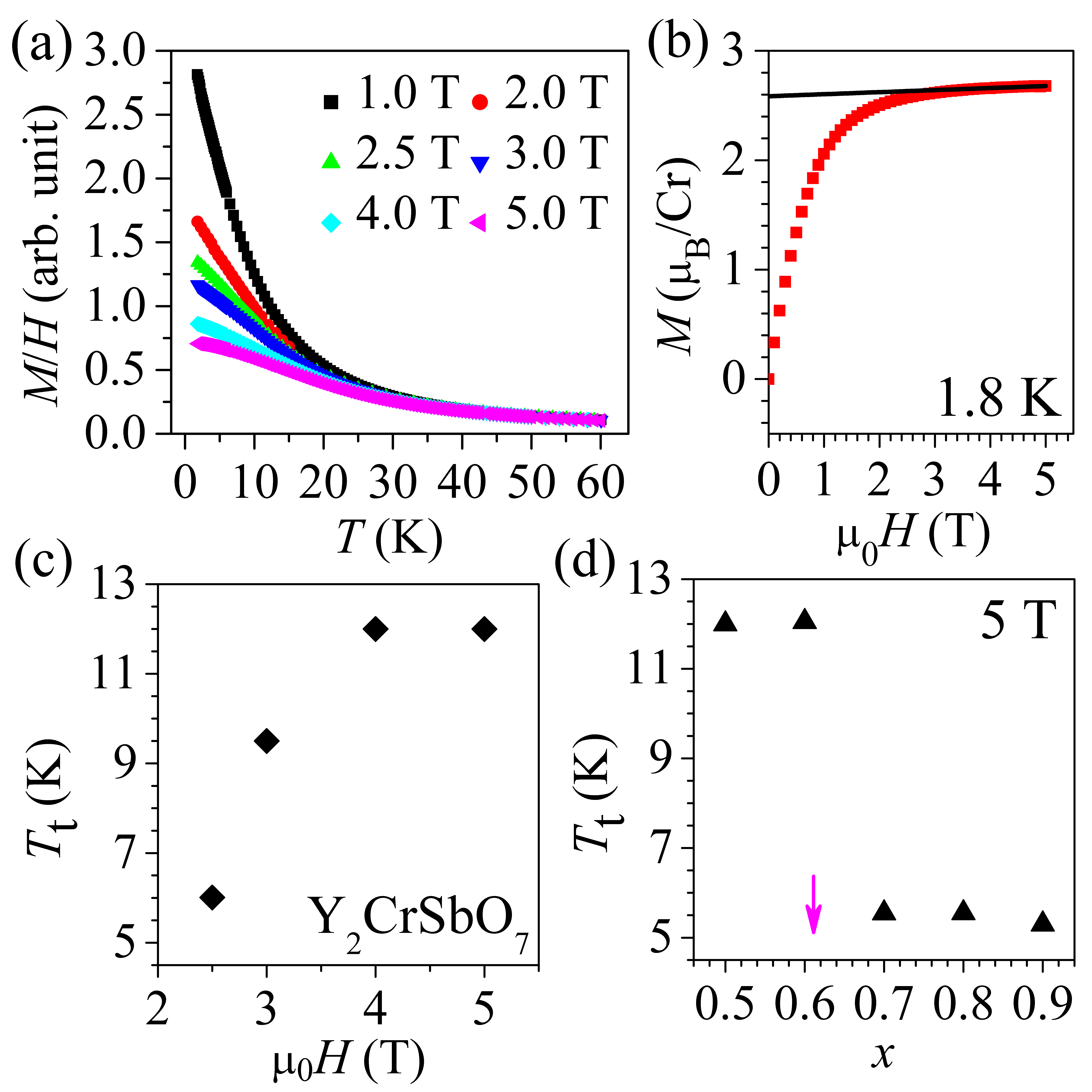}
	\caption{(a) Susceptibility (\textit{M}/\textit{H}) of Y$_{2}$CrSbO$_{7}$ versus temperature curves in the high field region. (b) Field scan of the magnetization of Y$_{2}$CrSbO$_{7}$ at 1.8\,K. The black line is a linear fit to the data above 3.5\,T. (c) Magnetic field dependence of the transition temperature ($T_\mathrm{{t}}$) in Y$_{2}$CrSbO$_{7}$. (d) \textit{x}-dependence of $T_\mathrm{{t}}$ measured at 5\,T, showing the recovery of site percolation. The vertical arrow marks the position of $x_{s}$.}	
	\label{fig:3}
\end{figure}

Within the resolution of our Rietveld refinement ($\sim$\,1\,$\%$), the Cr\,:\,Sb ratio is 1\,:\,1 in Y$_{2}$CrSbO$_{7}$.\ Based on Ref.~\onlinecite{Henley}, nonmagnetic sites percolate at $x_{s}$\ $\approx$\ 0.61 on a pyrochlore lattice.\ Thus, the Cr-sites in Y$_{2}$CrSbO$_{7}$ (\textit{x}\ =\ 0.5) are still percolative with a fraction $f_\mathrm{{M}}$\ =\ 83(2)\,$\%$ (see Fig.\,\ref{fig:2}c obtained from our simulations, details of which will be discussed below). For a ferromagnetic TM pyrochlore, magnetic frustration is often negligible due to the lack of possible sources. The expected ferromagnetically ordered moment ($M_\mathrm{{exp}}$) in Y$_{2}$CrSbO$_{7}$ can be estimated to be 0.83\,$\times$\,\textit{gJ}\,$\sim$ 2.59\,$\mu_{B}$/Cr, where the Land\'e \textit{g}-factor is approximately 2 and \textit{J}\ =\ 1.56 is extracted from $M_\mathrm{{eff}}$.\ $M_\mathrm{{exp}}$ is well above the resolution of our HRNPD measurements\ (<\ 0.5\,$\mu_{B}$). The magnetic frustration in Y$_{2}$CrSbO$_{7}$ is reflected by the frustration index, $\mathit{h}$\ =\ $|\dfrac{\Theta_\mathrm{{CW}}}{T_\mathrm{{t}}}|$\ >\ 10, where $T_\mathrm{{t}}$ is the transition temperature. $\mathit{h}$ is 2.7 in Y$_{2}$Mn$_{2}$O$_{7}$ and close to 1 in non-frustrated magnets. This high level of frustration is usually not expected for ferromagnetically coupled spins. One possible origin for the suppressed $T_\mathrm{{t}}$ in Y$_{2}$CrSbO$_{7}$ is nonmagnetic site-disorder. The critical concentration where the magnetic long range order disappears is very close, if not equal, to $x_{s}$\ $\approx$\ 0.61\,\cite{Cheong,Breed,Kumar,Tahir}, meaning conventional ferromagnetism should still develop in Y$_{2}$CrSbO$_{7}$ (\textit{x}\ =\ 0.5) below 1.8\,K. However, \textit{M-T} curves in the high field region do not support this scenario (Fig.\,\ref{fig:3}). These magnetic fields are expected to smooth out the ferromagnetic phase transition and leave $T_\mathrm{{t}}$ unchanged\,\cite{Blundell}.     

When diluting the magnetic sites by nonmagnetic ions, bond-disorder is also introduced to the local lattice due to the inevitable ionic size mismatch between \textit{M} and \textit{N}. As a result, there will be five types of \textit{M}/\textit{N}-tetrahedron in Y$_{2}$(\textit{M}$_{1-x}$\textit{N}$_{x}$)$_{2}$O$_{7}$, which can be labelled as empty, single, double, triple, and full in terms of \textit{M} occupation (Fig.\,\ref{fig:2}a). For site percolation, bond-disorder is ignored so that the percolation of doubly, triply, and fully occupied tetrahedra are calculated simultaneously. For the bond percolation, on the other hand, bond-disorder is the focus. Bond-disorder will randomly distort the local TMO$_{6}$-octahedron, frustrating the crystal field. Moreover, the random distribution of \textit{M}-O-\textit{M} bond angles caused by bond-disorder will lead to exchange fluctuations. Similar effects have been intensively studied in the SG pyrochlore Y$_{2}$Mo$_{2}$O$_{7}$, where the Mo-Mo exchange is antiferromagnetic and bond-disorder comes from orbital frustration\,\cite{Saunders,Shinaoka,Paddison,Greedan3}.

To qualitatively elucidate the effect of bond-disorder, we have simulated the percolation processes of Cr-sites with at least two (site percolation), three, four, five, and six (bond percolation) nearest neighbour Cr-sites, respectively. The simulations were performed on a \textit{D}\,$\times$\,\textit{D}\,$\times$\,\textit{D} (\textit{D}\ =\ 48, 64) cubic pyrochlore lattice (edge-sharing tetrahedral network). This lattice, initially with all sites occupied by magnetic ions (\textit{x}\ =\ 0), were randomly diluted by nonmagnetic ions to the required composition \textit{x}. For each composition, the percolation probability is 1 if at least one percolative path is found between any of the two parallel facets of the cube in our simulations. The simulations for each \textit{x} were sampled by \textit{s} times. As shown in Fig.\,\ref{fig:2}b, the site percolation threshold, $x_{s}$\ =\ 0.61(1), produced by our simulations is consistent with the previous study\,\cite{Henley}.\ In addition, our simulations also predict that bond percolation occurs well ahead of site percolation at $x_{b}$\ =\ 0.23(1) (Fig.\,\ref{fig:2}b). Due to the valence constraint, we are unable to check $x_{b}$ in our samples, Y$_{2}$(Cr$_{1-x}$Ga$_{x-0.5}$Sb$_{0.5}$)$_{2}$O$_{7}$ (0.5 $\leqslant$\,\textit{x}\,$\leqslant$\,0.9).\ However, we have identified a spinel system, Zn(Cr$_{1-x}$Ga$_{x}$)$_{2}$O$_{4}$, where the Cr/Ga-sites form a pyrochlore sublattice. The clean compound ZnCr$_{2}$O$_{4}$ (\textit{x}\ =\ 0) undergoes a spin-Peierls-like phase transition at $T_\mathrm{{N}}$\ =\ 12.5\,K\,\cite{Lee2}. By increasing the nonmagnetic Ga fraction (\textit{x}) on Cr-sites, the magnetically ordered moment drops rapidly to zero; a N\'eel to SG transition sets in between 0.2 and 0.25\,\cite{Lee}. These results are in excellent agreement with our bond percolation model (Fig.\,\ref{fig:2}). The transition temperatures of both N\'eel order and SG decrease monotonously as a function of \textit{x} in Zn(Cr$_{1-x}$Ga$_{x}$)$_{2}$O$_{4}$\,\cite{Lee}. This suggests that Y$_{2}$CrSbO$_{7}$ might undergo a transition at very low temperatures. Theoretically, bond-disorder is responsible for the onset of the SG state in Zn(Cr$_{1-x}$Ga$_{x}$)$_{2}$O$_{4}$\,\cite{Saunders,Shinaoka}. While the magnetic ground state of Y$_{2}$CrSbO$_{7}$ is not experimentally confirmed yet, it is likely that it is also a SG caused by bond-disorder.

The ionic radii of Cr$^{3+}$, Ga$^{3+}$, and Sb$^{5+}$ are 0.615\,\AA{}, 0.62\,\AA{}, and 0.60\,\AA{}, respectively\,\cite{Shannon}.\ As a result, the strength of bond-disorder in this system is very weak, but still sufficient to see an effect. Advanced SG theories, which can be applied to Zn(Cr$_{1-x}$Ga$_{x}$)$_{2}$O$_{4}$ and Y$_{2}$Mo$_{2}$O$_{7}$, have demonstrated a spin freezing transition in the zero bond-disorder limit.\ To estimate the strength of bond-disorder in Y$_{2}$(Cr$_{1-x}$Ga$_{x-0.5}$Sb$_{0.5}$)$_{2}$O$_{7}$, we examine the magnetic properties under an external perturbation, i.e. magnetic field. As shown in Fig.\,\ref{fig:3}a, a low temperature magnetization plateau gradually develops in Y$_{2}$CrSbO$_{7}$ as magnetic field is increased. The field-dependence of the magnetization is displayed in Fig.\,\ref{fig:3}b. The magnetization is saturated between 3.0 and 4.0\,T with $M_\mathrm{{s}}$\ =\ 2.59\,$\mu_{B}$/Cr\ =\ $M_\mathrm{{exp}}$. This may suggest that only the spins on percolative Cr-sites are ordered. The transition temperature $T_\mathrm{{t}}$ is defined as the point where the corresponding \textit{M}-\textit{T} curve has the steepest slope. The continuous increase of $T_\mathrm{{t}}$ as a function of magnetic field ($\mu_\mathrm{{0}}$H\ <\ 4\,T) supports the idea of a highly frustrated zero-field magnetic ground state with nonuniform bond-disorder. At 4\,T and above, $T_\mathrm{{t}}$ of Y$_{2}$CrSbO$_{7}$ tends to saturate $\sim$\ 12\,K (Fig.\,\ref{fig:3}c). This deviates from the behaviour of a polarized paramagnet\,\cite{Blundell}. We note that $T_\mathrm{{t}}$ only saturates when the percolative Cr-spins are fully aligned. In other words, magnetic long range order is fully recovered in Y$_{2}$CrSbO$_{7}$ when the magnetic frustration is removed. As a result, the strength of bond-disorder in Y$_{2}$CrSbO$_{7}$ is estimated to range from 0\,T and 3.5\,T. A sudden drop in $T_\mathrm{{t}}$ measured at 5\,T is observed between 0.6 and 0.7 (Fig.\,\ref{fig:3}d).\ It means that long range spin correlation, which is recovered by suppressing the magnetic frustration caused by bond-disorder, develops in the high field region for \textit{x}\ <\ $x_{s}$\ =\ 0.61(1). It also indicates that our samples are stoichiometrically homogeneous with $\Delta$\textit{x}\ <\ 0.02.
\section{Summary}
Based on our magnetization and HRNPD measurements (Fig.\,\ref{fig:1}), we have observed a very high level of magnetic frustration ($\mathit{h}$\ >\ 10) in the TM pyrochlore Y$_{2}$CrSbO$_{7}$ where the Cr-Cr exchange is predominantly ferromagnetic. The magnetic frustration cannot be explained by nonmagnetic site-disorder (Sb). We propose percolative bond-disorder caused by the ionic size mismatch as the driving mechanism. The average Cr/Sb-O-Cr/Sb bond angle is 130.19(1) degrees in Y$_{2}$CrSbO$_{7}$. Based on a previous study on the Cr-based oxides with very similar lattice parameters, this value is in the critical region where the Cr-Cr exchange interaction changes its sign\,\cite{Motida}. Because of this, zero point exchange fluctuations might be present although the overall exchange is ferromagnetic. Secondly, bond-disorder will also affect the local crystal field environment, e.g. frustrating the single-ion anisotropy. We have also estimated the strength of bond-disorder in Y$_{2}$CrSbO$_{7}$, which is in the region of [0\,T,\ 3.5\,T]. As a result, both magnetic long range order and site percolation process can be recovered by applying high magnetic fields (Fig.\,\ref{fig:3}).

For Y$_{2}$(\textit{M}$_{1-x}$\textit{N}$_{x}$)$_{2}$O$_{7}$ (\textit{M}\ =\ magnetic TM ion, \textit{N}\ =\ nonmagnetic ion), we have performed numerical simulations to study the percolation process of various clusters, including bond-, site-disorder, and several intermediate states in between. Our results unambiguously reveal that bond-disorder [$x_{b}$\ =\ 0.23(1)] percolates well ahead of site-disorder [$x_{s}$\ =\ 0.61(1)]. More importantly, our model can be experimentally verified in the spinel system Zn(Cr$_{1-x}$Ga$_{x}$)$_{2}$O$_{4}$ where a N\'eel to SG transition can be induced by Ga-substitution (Fig.\,\ref{fig:2})\,\cite{Lee}. Considering the similarity between the Cr/Sb- and Cr/Ga- sublattices in many aspects, we propose that the magnetic ground state of Y$_{2}$CrSbO$_{7}$ is a SG. We also call for further investigations in future. For example, experiments, e.g. heat capacity and HRNPD, in the ultra-low temperature region (<\ 1.8\,K) would be helpful to understand the magnetic ground state of Y$_{2}$CrSbO$_{7}$. Investigations on the local crystallographic structure are needed to check the amplitude of exchange fluctuations.

\begin{acknowledgments}
We thank M. W. Long and E. M. Forgan for helpful discussions. We acknowledge the UK EPSRC for funding under grant number EP/J016977/1.
\end{acknowledgments}

\end{document}